\documentclass[a4paper]{jpconf}

\usepackage{graphicx}% Include figure files
\usepackage{dcolumn}% Align table columns on decimal point
\usepackage{bm}% bold math
\usepackage{hyperref}% add hypertext capabilities
\usepackage{color}
\usepackage{breakurl}
\usepackage{amsmath}

\newcommand{\signal}{\mathcal{S}}
\newcommand{\bg}{\mathcal{B}}

\begin{document}

\begin{flushright}
  MPP-2011-142
\end{flushright}

\title{Spectral cutoffs in indirect dark matter searches}

\author{Christoph Weniger$^\ast$}
\address{Max-Planck-Institut f\"ur Physik, F\"ohringer Ring 6, 80805
Munich, Germany}
\ead{weniger@mppmu.mpg.de}
\author{Torsten Bringmann$^1$, Francesca Calore$^1$ and Gilles Vertongen$^2$}
\address{$^1$ {II.} Institute for Theoretical Physics, University of Hamburg, Luruper Chaussee 149, DE-22761 Hamburg, Germany}
\address{$^2$ Institut d'Astrophysique de Paris, UMR-7095 du CNRS, 98 bis bd
Arago, 75014 Paris, France}
\address{$^\ast$ presenting author}

\begin{abstract}
  Indirect searches for dark matter annihilation or decay products in the
  cosmic-ray spectrum are plagued by the question of how to disentangle a dark
  matter signal from the omnipresent astrophysical background. One of the
  practically background-free smoking-gun signatures for dark matter would be
  the observation of a sharp cutoff or a pronounced bump in the gamma-ray
  energy spectrum. Such features are generically produced in many dark matter
  models by internal Bremsstrahlung, and they can be treated in a similar
  manner as the traditionally looked-for gamma-ray lines. Here, we discuss
  prospects for seeing such features with present and future Atmospheric
  Cherenkov Telescopes.
\end{abstract}

\section{Introduction}
Indirect dark matter (DM) searches aim at seeing an excess in cosmic rays from
the annihilation or decay of DM in the Galactic halo~\cite{Bertone:2004pz}.
Very often, indirect searches focus on \emph{secondary photons} from the
fragmentation of hadronic annihilation products.  The corresponding spectra
are rather broad and peak at energies much lower than the DM mass $m_\chi$,
which generically makes a convincing claim of a DM detection above the
astrophysical backgrounds difficult. In many models, however, \emph{pronounced
spectral features} are expected at the kinematic endpoint $E_\gamma=m_\chi$;
they include monochromatic gamma-ray lines \cite{Bergstrom:1997fj}, sharp
steps or cutoffs \cite{Birkedal:2005ep,Bergstrom:2004cy} as well as
pronounced bumps \cite{Bringmann:2007nk}. The type and strength of these
features are intricately linked to the particle nature of DM; a detection
would thus not only allow a convincing discrimination from astrophysical
backgrounds but also to determine important DM model parameters like the value
of $m_\chi$. So far, only line-signals have explicitly been searched
for~\cite{line_searches}---despite the fact that they are loop-suppressed and
thus generically subdominant compared to other spectral signatures
\cite{Bringmann:2007nk}. Here, we discuss a general method to search for sharp
spectral features in gamma-ray observations. Concentrating on DM models with a
large internal Bremsstrahlung (IB) component, and on observations with Imaging
Atmospheric Cherenkov Telescopes (IACTs), we derive projected limits and
prospects to see such signatures with current and future instruments.

\section{Methods and Targets}
\label{sec:method}
The defining aspect of the above-mentioned spectral features is an abrupt
change of the gamma-ray flux as function of energy.  It is therefore possible
to concentrate the search for spectral features on a small sliding energy
window $[E_0, E_1]$, with $E_0\!<\!  m_\chi\! <\! E_1$, and window sizes of
the order of a few times the energy resolution of the instrument. An important
advantage of considering only small window sizes is that astrophysical
gamma-ray fluxes can be typically very well described by a simple power-law.
Limits on and the significance for dark matter induced fluxes can then be
derived by standard statistical methods like the profile likelihood method. 

We are interested in deriving projected limits on the dark matter annihilation
cross-section that follow from the spectral end-point features of different
dark matter scenarios \textit{alone}, \textit{i.e.}~without taking into
account secondary photons, using the above method. We will focus on
observations of the Galactic center region with IACTs, considering benchmark
scenarios that roughly correspond to the telescope characteristics of the
currently operating H.E.S.S.~(IACT1)~\cite{Aharonian:2006pe}, the future CTA
(IACT2)~\cite{Consortium:2010bc} and---as the most optimistic choice for
indirect DM searches---the proposed Dark Matter Array (DMA,
IACT3)~\cite{Bergstrom:2010gh}.  For the background, we take into account
cosmic-ray fluxes of electrons and protons, the diffuse gamma-ray flux and the
source HESS J1745-290 at (or very close to) the Galactic center. We adopt an
Einasto dark matter profile, and a relatively small target region
$\Delta\Omega=2^\circ\times2^\circ$ around the Galactic center. The energy
window size is chosen such that the impact of a nonzero background curvature
on our final results is less than $50\%$ (see Ref.~\cite{us-old} for details).

\begin{table}[t]
  \begin{center}
    \begin{tabular}{lrrrrr}
      \br
      & DM particle  & $m_\chi^{\rm th}$ & $\langle\sigma v\rangle^{\rm th}$ & relevant & spectral\\
      && [TeV] & [cm$^3$s$^{-1}$] & channel & feature\\\mr
      $\gamma\gamma$  &  any WIMP & $\mathcal{O}$(0.1--10) & $\mathcal{O}(10^{-30})$& $\gamma\gamma$ & line\\
      KK  & $B^{(1)}$  & 1.3 & $1\cdot10^{-26}$ & $\ell^+\ell^-\gamma$ & FSR step\\
      BM3 &  neutralino & 0.23 & $9\cdot10^{-29}$ & $\ell^+\ell^-\gamma$ & IB bump\\
      BM4 &  neutralino & 1.9 & $3\cdot10^{-27}$ & $W^+W^-\gamma$ & IB bump\\\br
    \end{tabular}
  \end{center}
  \caption{DM benchmark models used in our analysis as examples for the
  typical spectral endpoint features to be expected in WIMP annihilations. For
  these particular models, we also state the annihilation channel that is most
  important in this context, as well as  mass and total annihilation rate for
  thermally produced DM. See text and Ref.~\cite{us-old} for further details
  about the DM models and Fig.~\ref{fig:DMspectra} for the corresponding
  photon spectra.}
  \label{tab:DM}
\end{table}

\begin{figure}[t]
  \includegraphics[width=0.50\linewidth]{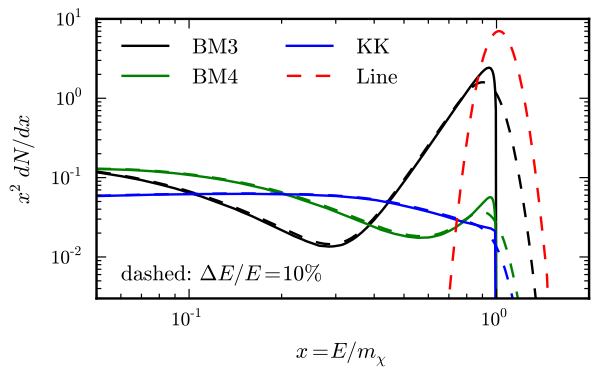}
  \hspace{0.1in}
  \begin{minipage}[b]{2.8in}
    \caption{Photon spectra for the DM benchmark models of Tab.~\ref{tab:DM}.
    Dashed lines show the same spectra smeared with a Gaussian of width
    $\Delta x/x=0.1$. From Ref.~\cite{us-old}.}
    \vspace{0.35in}
    \label{fig:DMspectra} 
  \end{minipage}
\end{figure}

\smallskip

We will discuss three types of typical endpoint features that arise from
radiative corrections to the tree-level annihilation process: (1) The most
striking spectral signature, in terms of a possible discrimination from a
power-law background, is a \emph{gamma-ray line}, which would result from the
direct annihilation of DM into $\gamma \gamma$, $Z \gamma$ or $H\gamma$.
Generically, the annihilation rate is loop-suppressed and expected to be of
the order of $\langle\sigma v\rangle_{\rm line}\sim\alpha_{\rm
em}^2\times\langle\sigma v\rangle_{\rm tree}\sim10^{-30}{\rm cm}^3{\rm
s}^{-1}$ (although in some cases much stronger line signals are possible).
(2) As an example for a  step-like feature we use the gamma-ray spectrum
expected from annihilating \emph{Kaluza-Klein (KK) DM} in models of universal
extra dimensions~\cite{Servant:2002aq}. Its total gamma-ray annihilation
spectrum at high energies is dominated by final state radiation (FSR) off
lepton final states, and the shape of the spectrum $dN/dx$, with
$x=E_\gamma/m_\chi$, turns out to be essentially independent of $m_\chi$ and
other model parameters~\cite{Bergstrom:2004cy}. (3) Pronounced bump-like
features at $E \simeq m_\chi$ may arise from IB in the annihilation of
\emph{neutralino DM} \cite{Bringmann:2007nk}.  Here, BM3 is a typical example
for a neutralino in the stau co-annihilation region, where photon emission
from virtual sleptons greatly enhances the photon spectrum at high energies;
BM4 refers to a situation in which IB from $W^\pm$ final states dominates. 

\begin{figure}[t]
  \includegraphics[width=\linewidth]{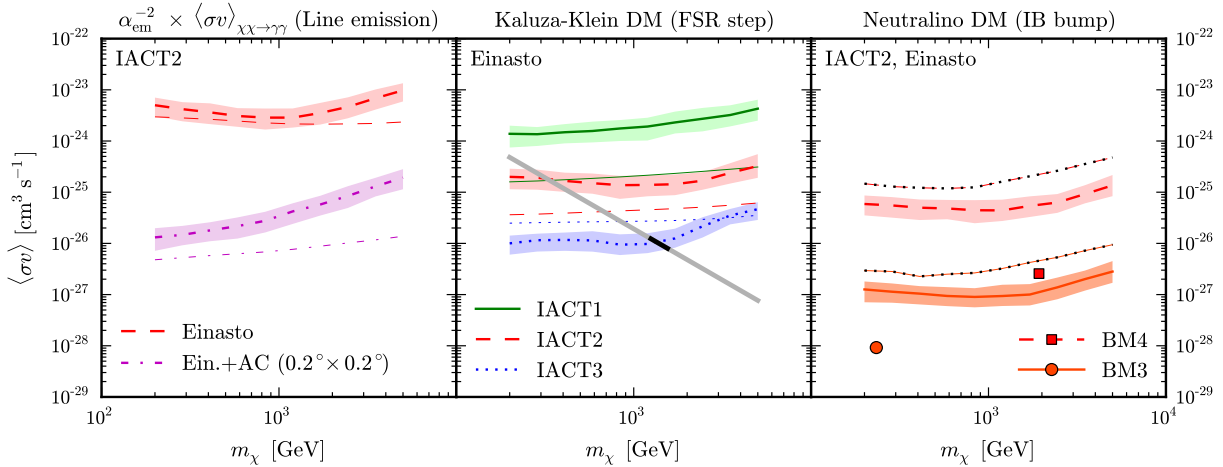}
  \caption{\textit{Thick lines:} Expected $2\sigma$ upper limits on
  $\langle\sigma v\rangle$ for  selected DM models, DM profiles and
  observational scenarios; bands indicate the variance of these limits.
  \textit{Thin lines:} Spectral feature of DM signal has
  $\signal/\bg\approx1\%$ (after convolution with energy dispersion).  The
  \textit{left panel} shows limits on gamma-ray lines, rescaled by a
  loop-factor of $\alpha_\text{em}^{-2}$ for better comparison.  In the
  \textit{central panel}, the gray band indicates the expected $\langle\sigma
  v\rangle$ for KK DM, the black part being compatible with the observed relic
  density. In the \textit{right panel}, we indicate the adopted neutralino
  benchmark points, and the \textit{dotted lines} show the projected $5\sigma$
  sensitivity. The benchmark experiments IACT1, IACT2 and IACT3 correspond
  roughly to H.E.S.S., CTA and DMA, respectively, see text
  for details. The figure is taken from Ref.~\cite{us-old}.}
  \label{fig:CS} 
\end{figure}

In Tab.~\ref{tab:DM}, we shortly summarize the properties of the DM benchmark
models described above, including for completeness the actual DM mass and
total annihilation rate needed to obtain the observed relic density for
thermally produced DM. Note, however, that we essentially treat these values
as free parameters in our analysis and that we are rather interested in the
spectral shape of the annihilation signal, represented by $dN/dx$; in
Fig.~\ref{fig:DMspectra} we show these spectra for a direct comparison.

\section{Results and Discussion} 

In Fig.~\ref{fig:CS} we show our results for the expected $2\sigma$ upper
limits (thick lines) on the above DM models as well as the variance of these
limits among the mock data sets.  We find that in particular IB features in
the spectrum (right panel) have the potential to constrain the annihilation
rate at least down to values typically expected for thermal production,
$\langle \sigma v\rangle\sim3\cdot10^{-26}{\rm cm}^3{\rm s}^{-1}$, already for
modest assumptions about the DM distribution.  For models with very large IB
contributions like BM3, we find that our method would provide even stronger
limits on $\langle\sigma v\rangle$ than what was obtained by the HESS analysis
of the Galactic center region assuming annihilation into $\bar b
b$~\cite{Abramowski:2011hc}. 

For the case of not too strongly pronounced endpoint features (like line
signals in most models or the step for Kaluza-Klein DM), secondary photons
will usually be more powerful in constraining the total annihilation rate
$\langle\sigma v\rangle$.  However, in case of an adiabatically compressed
profile our limits could improve by maybe two orders of magnitude, as
demonstrated for gamma-ray lines in the left panel. As shown in the central
panel of Fig.~\ref{fig:CS}, the future CTA should be able to place limits
about one order of magnitude stronger than currently possible, and the
proposed DMA could further improve these by another factor of ten.

Finally, we would like to stress that our limits in general provide rather
\emph{complementary information} on the DM nature and can thus usually not
easily be compared with limits on secondary photons. In any case, from the
point of view of indirect DM searches, the \emph{detection} of the kinematic
cutoff will be much more interesting than the detection of secondary photons,
because they would provide rather unambiguous evidence for the DM nature of
the signal as well as allow to determine important parameters like the DM
mass. An obvious extension of the approach presented here is to apply it to
the \emph{discrimination} of models \cite{us}.

\ack
T.B.~and C.W.~thank the organizers of the TAUP 2011 for an interesting and
stimulating conference. T.B.~and F.C.~acknowledge support from the German
Research Foundation (DFG) through Emmy Noether grant BR 3954/1-1.


\bigskip

\begin{thebibliography}{99}
\bibitem{Bertone:2004pz}
  G.~Bertone, D.~Hooper, J.~Silk,
  %``Particle dark matter: Evidence, candidates and constraints,''
  Phys.\ Rept.\  {\bf 405} (2005)  279.
%  [hep-ph/0404175].

\bibitem{Bergstrom:1997fj}
  L.~Bergstr\"om, P.~Ullio and J.~H.~Buckley,
  %``Observability of gamma rays from dark matter neutralino annihilations  in
  %the Milky Way halo,''
  Astropart.\ Phys.\  {\bf 9} (1998) 137
  [arXiv:astro-ph/9712318].
  %%CITATION = APHYE,9,137;%%

\bibitem{Birkedal:2005ep}
  A.~Birkedal,
  K.~T.~Matchev, M.~Perelstein and A.~Spray,
  %``Robust gamma ray signature of WIMP dark matter,''
  arXiv:hep-ph/0507194.
  %%CITATION = HEP-PH/0507194;%%

\bibitem{Bergstrom:2004cy}
  L.~Bergstr\"om,
  T.~Bringmann, M.~Eriksson and M.~Gustafsson,
  %``Gamma rays from Kaluza-Klein dark matter,''
  Phys.\ Rev.\ Lett.\  {\bf 94} (2005) 131301
  [arXiv:astro-ph/0410359].
  %%CITATION = PRLTA,94,131301;%%

\bibitem{Bringmann:2007nk}
  T.~Bringmann, L.~Bergstr\"om and J.~Edsj\"o,
  %``New Gamma-Ray Contributions to Supersymmetric Dark Matter Annihilation,''
  JHEP {\bf 0801} (2008) 049
  [arXiv:0710.3169 [hep-ph]].
  %%CITATION = JHEPA,0801,049;%%

\bibitem{line_searches}
  \mbox{A.A.~Pullen, R.R.~Chary, M.~Kamionkowski,
  %``Search with EGRET for a gamma ray line from the galactic center,''
  Phys.}~Rev. D {\bf 76} (2007) 063006
  [Erratum-ibid.\  {\bf 83} (2011) 029904]
 [arXiv:astro-ph/0610295];
  %%CITATION = PHRVA,D76,063006;%%
  F.~A.~Aharonian {\it et al.} [ HEGRA Collaboration ],
  %``Search for TeV gamma-ray emission from the Andromeda galaxy,''
  Astron.\ Astrophys.\  {\bf 400}, 153 (2003)
  [astro-ph/0302347];  
  A.~A.~Abdo {\it et al.},
  %``Fermi LAT Search for Photon Lines from 30 to 200 GeV and Dark Matter
  %Implications,''
  Phys.\ Rev.\ Lett.\  {\bf 104} (2010) 091302
  [arXiv:1001.4836 [astro-ph.HE]];
  %%CITATION = PRLTA,104,091302;%%
  G.~Vertongen and C.~Weniger,
  %``Hunting Dark Matter Gamma-Ray Lines with the Fermi LAT,''
  JCAP {\bf 1105} (2011) 027
  [arXiv:1101.2610 [hep-ph]].
  %%CITATION = ARXIV:1101.2610;%%

\bibitem{Aharonian:2006pe}
  F.~Aharonian {\it et al.}
    [H.E.S.S. Collaboration],
  %``Observations of the Crab Nebula with H.E.S.S,''
  Astron.\ Astrophys.\  {\bf 457} (2006) 899 
  [arXiv:astro-ph/0607333].
% (from Fig.~13a, $20^\circ$).
  %%CITATION = AAEJA,457,899;%%

\bibitem{Consortium:2010bc}
  The CTA Consortium,
  %``Design Concepts for the Cherenkov Telescope Array,''
  arXiv:1008.3703 [astro-ph.IM].
  %%CITATION = ARXIV:1008.3703;%%

\bibitem{Bergstrom:2010gh}
  L.~Bergstr\"om, T.~Bringmann and J.~Edsj\"o,
  %``Complementarity of direct dark matter detection and indirect detection
  %through gamma-rays,''
  Phys.\ Rev.\  D {\bf 83} (2011) 045024
  [arXiv:1011.4514 [hep-ph]].
  %%CITATION = PHRVA,D83,045024;%%

\bibitem{us-old}
  T.~Bringmann, F.~Calore, G.~Vertongen and C.~Weniger,
  %``On the Relevance of Sharp Gamma-Ray Features for Indirect Dark
  %Matter Searches,''
  Phys.\ Rev.\ D {\bf 84} (2011) 103525
  [arXiv:1106.1874 [hep-ph]].
  %%CITATION = ARXIV:1106.1874;%%

\bibitem{Servant:2002aq}
  G.~Servant and T.~M.~P.~Tait,
  %``Is the lightest Kaluza-Klein particle a viable dark matter candidate?,''
  Nucl.\ Phys.\  B {\bf 650} (2003) 391
  [arXiv:hep-ph/0206071].
  %%CITATION = NUPHA,B650,391;%%

\bibitem{Abramowski:2011hc}
  A.~Abramowski {\it et al.},
  % [ H.E.S.S. Collaboration ],
  %``Search for a Dark Matter annihilation signal from the Galactic Center halo with H.E.S.S,''
  Phys.\ Rev.\ Lett.\  {\bf 106}, 161301 (2011)
  [arXiv:1103.3266 [astro-ph.HE]].

\bibitem{us}
  T.~Bringmann, F.~Calore, G.~Vertongen and C.~Weniger,
  work in progress; see also, \textit{e.g.},
  %\cite{arXiv:1007.0018}
  %\bibitem{arXiv:1007.0018}
  M.~Perelstein and B.~Shakya,
  %``Dark Matter Identification with Gamma Rays from Dwarf Galaxies,''
  JCAP\ {\bf 1010} (2010) 016
  [arXiv:1007.0018 [astro-ph.HE]].
  %%CITATION = JCAPA,1010,016;%%
\end{thebibliography}
\end{document}